\documentclass[letterpaper,english,prd,preprint]{revtex4-1}
\usepackage[LGR,T1]{fontenc}
\usepackage[latin9]{inputenc}
\usepackage{color}
\usepackage{babel}
\usepackage{array}
\usepackage{float}
\usepackage{multirow}
\usepackage{graphicx}
\usepackage{amsmath,xparse}
\usepackage[unicode=true,pdfusetitle,
 bookmarks=true,bookmarksnumbered=false,bookmarksopen=false,
 breaklinks=false,pdfborder={0 0 0},pdfborderstyle={},backref=false,colorlinks=true]
 {hyperref}
\usepackage{breakurl}
\usepackage{tikz}
\usetikzlibrary{patterns}
\usetikzlibrary{plotmarks}

\makeatletter

\ProvideTextCommand{\~}{LGR}[1]{\char126#1}

\providecommand{\tabularnewline}{\\}

\@ifundefined{date}{}{\date{}}
\usepackage[compat=1.1.0]{tikz-feynman}
\usepackage{feynmp}
\usepackage{pgf,tikz}
\usepackage{color}
\usepackage{upgreek}
\usepackage{hepnames}
\usetikzlibrary{trees}
\usetikzlibrary{positioning,arrows}
\usetikzlibrary{decorations.pathmorphing}
\usetikzlibrary{decorations.markings}
\usepackage{pgfplots}
\usepackage[none]{hyphenat}
\pgfplotsset{compat=newest}
\tikzset{
photon/.style={decorate, decoration={snake}, draw=red},
particle/.style={draw=blue, postaction={decorate},
decoration={markings,mark=at position .5 with {\arrow[draw=blue]{>}}}},
antiparticle/.style={draw=blue, postaction={decorate},
decoration={markings,mark=at position .5 with {\arrow[draw=blue]{<}}}},
gluon/.style={decorate, draw=black,
decoration={coil,amplitude=4pt, segment length=5pt}}
}
\makeatother

\begin{document}


\title{Search for Excited $u$ and $d$ Quarks in Dijet Final States at Future $pp$ Colliders}

\author{Ahmet Nuri Akay}
\email{ahmetnakay@gmail.com}

\selectlanguage{english}%

\affiliation{TOBB Economics and Technology University, Ankara/TURKEY}

\author{Yusuf Oguzhan Günaydin}
\email{yusufgunaydin@gmail.com}

\selectlanguage{english}%

\affiliation{Department of Physics, Kahramanmaras Sütcü Imam University, Kahramanmaras/TURKEY}

\author{Mehmet Sahin}
\email{mehmet.sahin@usak.edu.tr}

\selectlanguage{english}%

\affiliation{Department of Physics, Usak University, Usak/TURKEY }

\author{Saleh Sultansoy}
\email{ssultansoy@etu.edu.tr}

\selectlanguage{english}%

\affiliation{TOBB Economics and Technology University, Ankara/TURKEY}

\affiliation{ANAS, Institute of Physics, Baku/AZERBAIJAN}

\date{\today}
\begin{abstract}
Resonant production of excited $u$ and $d$ quarks at the Future Circular Collider
and Super proton-proton Collider have been researched. Dominant jet-jet
decay mode has been considered. It is shown that FCC and SppC have
great potential for discovery of excited $u$ ($d$) quark: up to 44.1 (36.3) and 58.4 (47.8)
TeV masses, respectively. For degenerate case (M$_{u^{\star}}$ = M$_{d^{\star}}$), these values are 45.9 and 60.9 TeV, respectively. This discovery will also afford an opportunity to determine the compositeness scale up to multi-PeV level. 
\end{abstract}



\maketitle

\section{\label{sec:intro}INTRODUCTION}
Standard Model (SM) contains plenty of elementary particles and their
parameters that are not completely explained. To overcome these unsolved
problems that the SM does not give answers, new models have been developed
beyond the Standard Model (BSM) such as composite models, supersymmetry, extra dimensions,
string theory, and so on. These BSM theories require higher
energy level than SM energy domain to bring solutions for unanswered
problems. Therefore, the SM is considered as low energy configuration of
the more fundamental theory. 

Numbers of particles and parameters in the SM are reduced in the frame
of the composite models \cite{pati1974,terazawa1977,shupe1979,harari1979,terazawa1980,fritzsch1981,terazawa1982,terazawa1983,eichten1983,dsouza1992,celikel1998,desouza2008,terazawa2014,terazawa2015,fritzsch2016,Kaya2018}.
According to composite models, while SM quarks and leptons are predicted
as composite particles, preons are considered as the most fundamental
particles. If excited states of the SM fermions are experimentally
observed, this observation will be clear proof of quarks and leptons'
compositeness. 

Excited fermions are known to represent much heavier particles than the
SM fermions and they could be split into two classes: excited quarks
($q^{\star}$) and excited leptons ($l^{\star}$). These heavy particles could
also have spin-1/2 and spin-3/2 states. From the first publication
on excited leptons in 1965 \cite{low1965} until today, there have
been plenty of phenomenological \cite{renard1983,lyons1983,kuhn1984,pancheri1984,rujula1984,hagiwara1985,kuhn1985,baur1987,spira1989,baur1990,jikia1990,boudjema1993,cakir1999,cakir2000,cakir2001,eboli2002,cakir2004,ccakir2004,cakir2008,caliskan2017,caliskan2017a}
and experimental \cite{cdf1995,h1_2000,l3_2000,chekanov2002,abdallah2006,cms2014,atlas2016photon,atlas2016,atlas2017,cms2016,cms2016photon,cms2017,sirunyan2018,sirunyan2018gammajet,sirunyan2018WjetZjet} studies performed on excited fermions.

Excited states of SM quarks might be shown in four possible final
states with light jets, $q^{\star}\rightarrow jj$, $q^{\star}\rightarrow\gamma j$,
$q^{\star}\rightarrow Wj$ and $q^{\star}\rightarrow Zj$. Currently, the
LHC puts experimental mass limits for all four final state cases \cite{pdg2016,atlas2017,sirunyan2018, sirunyan2018gammajet,sirunyan2018WjetZjet}
that are M$_{q^{\star}}=6.0$ ($ 6.0$), $5.5$ ($ 5.5$), $3.2$ ($ 5.0$), and $2.9$ ($4.7 $) TeV for ATLAS (CMS), respectively.
Like SM fermions, excited fermions also have three families and we
focused on $u^{\star}$ and $d^{\star}$  productions which decay to dijet final states. 

After the LHC physics mission is over, a new and more powerful collider
will take place as an energy frontier discovery machine for the high
energy physics. At CERN located in Geneva, Future Circular Collider (FCC) \cite{fcc2014}
is planned for the next step with $\sqrt{s}=100$ TeV. The other project,
Super Proton Proton Collider (SppC) is planned in China at multi TeV
center of mass (CM) energies \cite{su2016}, we chose $\sqrt{s}=136$
TeV option in this study. Both projects promise very high luminosity.
The FCC will be expected to reach 2500 $fb^{-1}$ integrated luminosity
in ten years (Phase I) and 15000 $fb^{-1}$ integrated luminosity
in 15 years (Phase II) \cite{benedikt2015fcc,benedikt2015fcclumi,benedikt2016towards}.
Overall in 25 years, total integrated luminosity will be 17500 $fb^{-1}$.
On the other hand, the SppC will deliver $pp$ collisions with 22500
$fb^{-1}$ integrated luminosity in 15 years. (See Tab. \ref{tab:FCC-and-SppC})

\begin{table}[H]
	\caption{Planned operation time of FCC and SppC and their main parameters\label{tab:FCC-and-SppC}}
	\begin{center}
		\begin{tabular}{|>{\raggedright}p{4cm}|>{\centering}p{3cm}|>{\centering}p{3cm}|>{\centering}p{2cm}|}
			\hline 
			\multirow{2}{4cm}{	\textbf{Collider Name} }& \multicolumn{2}{c|}{\textbf{FCC}} &\multirow{2}{2cm}{ \textbf{SppC}}\tabularnewline
			\cline{2-3}
					 & \textbf{Phase I} & \textbf{Phase II } & \tabularnewline
			\hline 
			
				\textbf{\textbf{Operation Time}} &  10 Years & 15 Years & 15 Years\tabularnewline
			\hline 
			
		\textbf{$\sqrt{s}$ [TeV]} & \multicolumn{2}{c|}{100} & 136\tabularnewline
			\hline 
			\multirow{2}{4cm}{\textbf{$\mathcal{L}_{int}$ [fb$^{-1}$}]}	& 2500  & 15000  & \multirow{2}{0.5cm}{22500 }\tabularnewline
			\cline{2-3} 
			& \multicolumn{2}{c|}{17500 } & \tabularnewline
			\hline 
		\end{tabular}
	\end{center}
\end{table}

In this research, we explore spin-1/2 excited $u$ and $d$ quark ($u^{\star}$ and $d^{\star}$) decaying into dijet final states at the FCC and the SppC. In the following sections, we state spin-1/2 excited quark interaction Lagrangian, decay widths and cross section values in section \ref{sec:Interaction-Lagrangian}, signal-background analysis
to determine cuts in section \ref{sec:Signal-BackGround-Analysis},  and attainable mass and compositeness scale ($\Lambda$) limits and conclusions in section \ref{sec:Results-and-Conclusions}. 

\section{\label{sec:Interaction-Lagrangian}Interaction Lagrangian, Decay Widths \protect\lowercase{and} Cross Sections}

When left- and right-handed components of excited quarks are assigned to isodoublets,  isospin structure of the first generation SM and excited quarks will be 
\[ \begin{bmatrix} 
u \\
d 
\end{bmatrix}_L , 
\begin{array}{c}
u_R, \; d_R 
\end{array}
\begin{bmatrix} 
u^{\star} \\
d^{\star} 
\end{bmatrix}_L , 
\begin{bmatrix} 
u^{\star} \\
d^{\star} 
\end{bmatrix}_R. \]

Since interaction Lagrangian is magnetic-moment type, it contains only left-handed quark doublet and consequently right-handed excited quark doublet. For that reason, as an effective interaction Lagrangian \cite{kuhn1984,rujula1984,baur1987,pdg2016},
Equation \ref{eq:lagrangian} was utilized for the spin-1/2 excited
quarks :

\begin{equation}
L_{eff}=\frac{1}{2\Lambda}\overline{q^{\star}_R}\;\sigma^{\mu\nu}[g_{s}f_{s}\frac{\lambda_{a}}{2}G_{\mu\nu}^{a}+gf\frac{\overrightarrow{\tau}}{2}\overrightarrow{W}_{\mu\nu}+g^{'}f^{'}\frac{Y}{2}B_{\mu\nu}]q_L+h.c.\label{eq:lagrangian}
\end{equation}
where, compositeness scale is represented as $\Lambda$, $q^{\star}_R$
denotes right-handed excited quark doublet, $q_L$ depicts ground state left-handed quark doublet, field strength
tensors are $G_{\mu\nu}^{a}$ for gluon, $\overrightarrow{W}_{\mu\nu}$
for SU(2), and $B_{\mu\nu}$ for U(1). $\lambda_{a}$, $\overrightarrow{\tau}$,
and $Y$ are color parameters for gluon-quark interaction, Pauli spin
matrices and weak hyper-charge, respectively. Gauge coupling constants
are $g_{s}$, $g$, and $g'$; and $f_{s},f,f'$ are free parameters
that are taken as equal to 1 in numerical calculations. In addition, mentioned interactions with Higgs Boson as well as mass mixing among quarks and excited quarks can be neglected since M$_{u^{\star}} \gg \eta \gg$ M$_u$ ($\eta$ is vacuum expectation value of Higgs field). Indeed, M$_{u^{\star}} >$ 6 TeV from the LHC data, $\eta \approx$  245 GeV and M$_u$ is in MeV region.

Interaction Lagrangian (Eq. \ref{eq:lagrangian}) was implemented into CalcHEP
\cite{calchep2013} software by using LanHEP interface \cite{LanHEP,lanhep2016}.
In our calculations, CTEQ6L1 \cite{pumplin2002,stump2003} parton
distribution function was used and factorizations and renormalization
scale were taken equal to $M_{q^{\star}}$. 

\begin{table}[H]
	\caption{Third component of isospins, charges, decay channels and widths  of up- and down-type excited quarks.\label{tab:decayChannels}}
	
	\begin{tabular}{|>{\centering}p{2cm}|>{\centering}p{2cm}|>{\centering}m{4cm}|m{8cm}|}
		\hline 
	\textbf{T$_3$}	&\textbf{Q }& \textbf{Decay Modes} & \textbf{Partial Decay Widths}  \\ 
		\hline 
	\multirow{4}{0cm}{$\dfrac{1}{2}$}	&\multirow{4}{0cm}{$\dfrac{2}{3}$}	& $u^{\star}\rightarrow dW^+$ &$\Gamma = \frac{1}{32\pi}g_{W}^{2}f_{W}^{2}\frac{M_{u^{\star}}^{3}}{\Lambda^{2}}\left(1-\frac{m_{W}^{2}}{M_{u^{\star}}^{2}}\right)^{2}\left(2+\frac{m_{W}^{2}}{M_{u^{\star}}^{2}}\right) $ \\ 
		\cline{3-4}
	&	&$u^{\star}\rightarrow uZ$  &  $\Gamma = \frac{1}{32\pi}g_{Z}^{2}f_{Z}^{2}\frac{M_{u^{\star}}^{3}}{\Lambda^{2}}\left(1-\frac{m_{Z}^{2}}{M_{u^{\star}}^{2}}\right)^{2}\left(2+\frac{m_{Z}^{2}}{M_{u^{\star}}^{2}}\right)$\\ 
		\cline{3-4}
	&	&$u^{\star}\rightarrow ug$  & $\Gamma = \frac{1}{3}\alpha_{s}f_{s}^{2}\frac{M_{u^{\star}}^{3}}{\Lambda^{2}}$ \\ 
		\cline{3-4}
	&	&$u^{\star}\rightarrow u\gamma$  & $\Gamma = \frac{1}{4}\alpha f_{\gamma}^{2}\frac{M_{u^{\star}}^{3}}{\Lambda^{2}}$ \\ 
		\hline 
	\multirow{4}{0cm}{$-\dfrac{1}{2}$}	&	\multirow{4}{0cm}{$-\dfrac{1}{3}$}	&$d^{\star}\rightarrow uW^-$  & $\Gamma = \frac{1}{32\pi}g_{W}^{2}f_{W}^{2}\frac{M_{d^{\star}}^{3}}{\Lambda^{2}}\left(1-\frac{m_{W}^{2}}{M_{d^{\star}}^{2}}\right)^{2}\left(2+\frac{m_{W}^{2}}{M_{d^{\star}}^{2}}\right)$\\ 
		\cline{3-4}
	&	&$d^{\star}\rightarrow dZ$  &  $\Gamma = \frac{1}{32\pi}g_{Z}^{2}f_{Z}^{2}\frac{M_{d^{\star}}^{3}}{\Lambda^{2}}\left(1-\frac{m_{Z}^{2}}{M_{d^{\star}}^{2}}\right)^{2}\left(2+\frac{m_{Z}^{2}}{M_{d^{\star}}^{2}}\right)$\\ 
	\cline{3-4}
	& &$d^{\star}\rightarrow dg$  & $\Gamma = \frac{1}{3}\alpha_{s}f_{s}^{2}\frac{M_{d^{\star}}^{3}}{\Lambda^{2}}$ \\ 
	\cline{3-4}
	& &$d^{\star}\rightarrow d\gamma$  & $\Gamma = \frac{1}{4}\alpha f_{\gamma}^{2}\frac{M_{d^{\star}}^{3}}{\Lambda^{2}}$ \\ 
		\hline 
	\end{tabular} 
\end{table}

\begin{figure}[H]
	\begin{center}
		
		\includegraphics[scale=0.55]{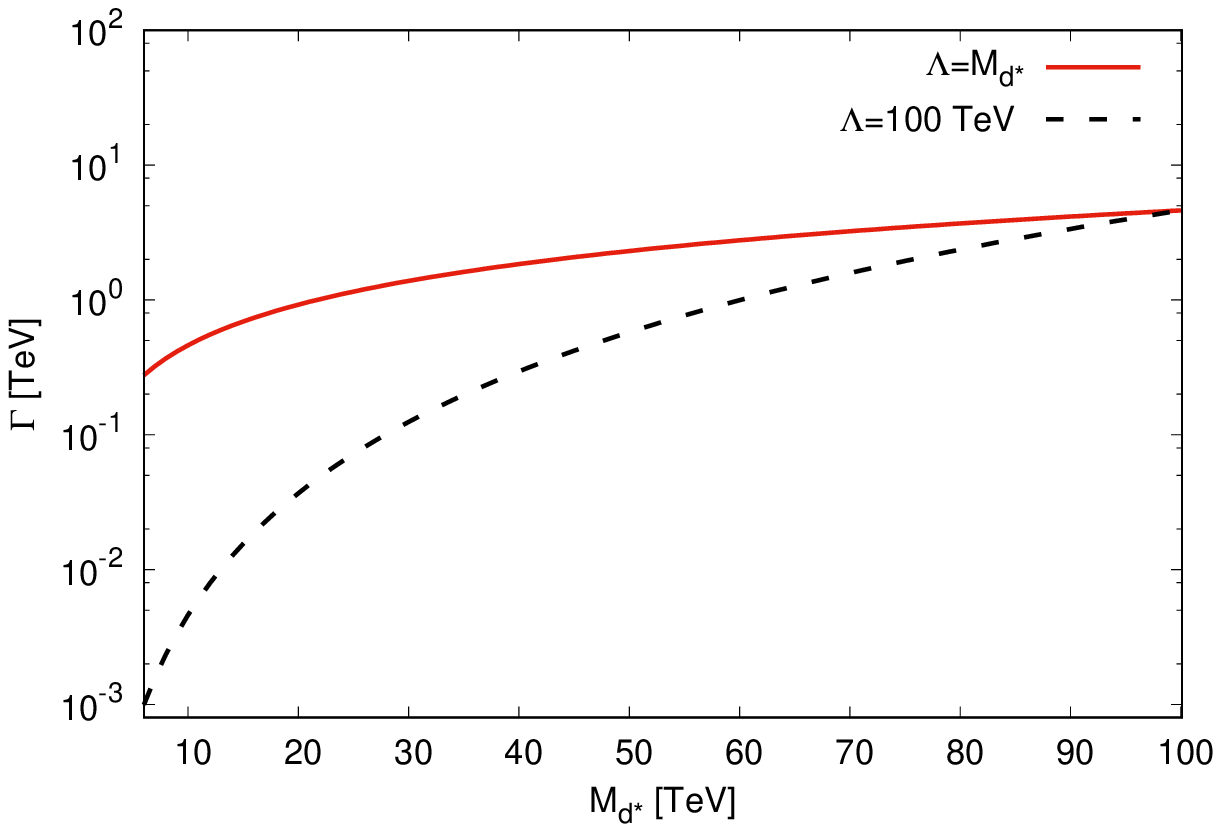}
		\includegraphics[scale=0.55]{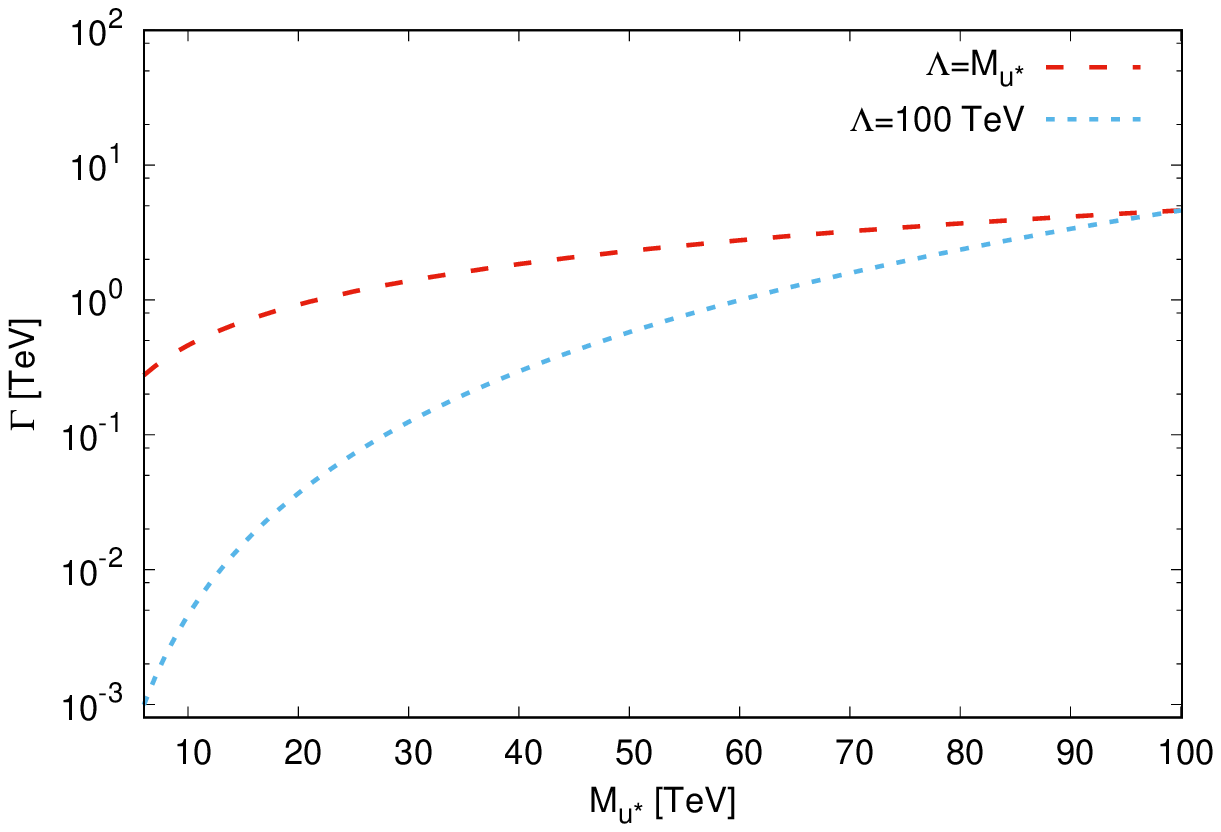}
		
		\caption{\label{fig:decayWidths}Decay widths versus first generation excited quark masses for both
			$\Lambda=M_{d^{\star}}$, $\Lambda=M_{u^{\star}}$ and $\Lambda=100$ TeV }
	\end{center}
\end{figure}

Partial decay widths of first generation excited quarks are listed in Tab. \ref{tab:decayChannels}. Parameters in the last column of Tab. \ref{tab:decayChannels} are  $f_{Z}=fT_{3}cos^{2}\theta_{W}-f^{'}(Y/2)sin^{2}\theta_{W}$,
$f_{W}=f/\sqrt{2}$, $f_{\gamma}=fT_{3}+f^{'}Y/2$, $g_{W}=\sqrt{4\pi\alpha}/sin\theta_{W}$,
and $g_{Z}=g_{W}/cos\theta_{W}$; here $T_{3}$ is the third component
of the weak isospin of  $q^{\star}$. In Fig. \ref{fig:decayWidths},
total decay widths are given for $\Lambda=M_{d^{\star}}$, $\Lambda=M_{u^{\star}}$ and $\Lambda=100$
TeV by scanning excited quarks mass from 6 TeV to 100 TeV. Total decay widths of $u^{\star}$ and $d^{\star}$ are close to each other since dominat decay modes are $u^{\star} \rightarrow g u$ and $d^{\star} \rightarrow g d$. There are small differences caused by $Z$ and $\gamma$ channels.  It is obviously
seen that while $d^{\star}$ and $u^{\star}$ mass values are risen, decay widths are increased.

\begin{center}	
	\begin{figure}[H]
		
		\includegraphics[scale=0.3]{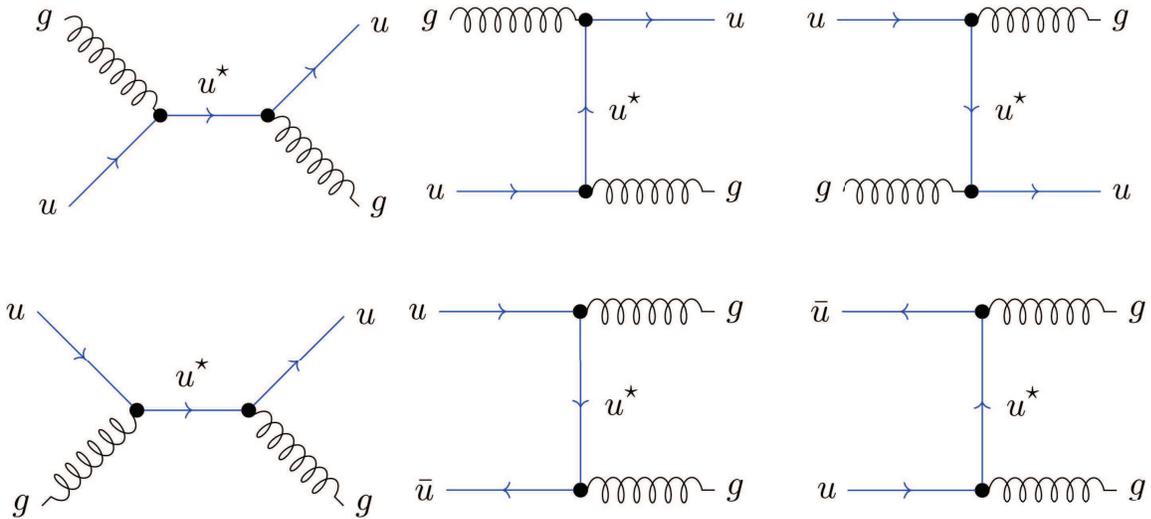}
		
		\caption{\label{fig:feynmanndiag} Feynman diagrams for direct (first column)
			and indirect production of $u^{\star}$ at $pp$ colliders. }
	\end{figure}
\end{center}

For the following parts of this study, we consider three cases to do analysis: \textbf{(a)} $M_{u^{\star}} < M_{d^{\star}}$, \textbf{(b)}$M_{d^{\star}} < M_{u^{\star}}$ and \textbf{(c)} $M_{u^{\star}} = M_{d^{\star}}$ (degenerate state) with $pp\rightarrow u^{\star}+X\rightarrow ug+X$, $pp\rightarrow d^{\star}+X\rightarrow dg+X$ and $pp\rightarrow q^{\star}+X\rightarrow qg+X$ signal processes, respectively (Here, $q^{\star}$ denotes $u^{\star} + d^{\star}$).
6 Feynman diagrams emerge for cases 
\textbf{(a) }and \textbf{(b)}, and 12 Feynman diagrams make contributions to signal cross section calculations for the case \textbf{(c)}.  Figure \ref{fig:feynmanndiag} presents the case \textbf{(a)} Feynman diagrams for illustration.  Analytical expression for the cross sections at parton level corresponding to these diagrams is described by equation \ref{eq:analytic}.

\begin{equation}
\begin{split}	
\dfrac{d\hat{\sigma}}{d\hat{t}} &= \dfrac{f_s^4 g_s^4}{216\pi\Lambda^4(M_{u^{\star}}^2-\hat{s})^2} \Bigg[\dfrac{-48M_{u^{\star}}^8+68M_{u^{\star}}^6\hat{s}+11M_{u^{\star}}^4\hat{s}^2-34M_{u^{\star}} ^2\hat{s}^3+6\hat{s}^4 }{(M_{u^{\star}}^2-\hat{s})^2 }  \\ 
&\hspace{4cm}+\dfrac{-8M_{u^{\star}} ^8-11M_{u^{\star}} ^6\hat{s}}{(M_{u^{\star}} ^2-\hat{t})^2} +\dfrac{32M_{u^{\star}} ^6+33M_{u^{\star}} ^4\hat{s}}{M_{u^{\star}} ^2-\hat{t}} -16\hat{s}\hat{t}\\
&\hspace{4cm}-16\hat{t}^2  +\dfrac{-8M_{u^{\star}}^8-11M_{u^{\star}}^6\hat{s}}{(M_{u^{\star}}^2+\hat{s}+\hat{t})^2}+\dfrac{32M_{u^{\star}}^6+33M_{u^{\star}}^4\hat{s}}{M_{u^{\star}}^2+\hat{s}+\hat{t}}\Bigg]
\end{split}
\label{eq:analytic}
\end{equation}

\begin{figure}[H]
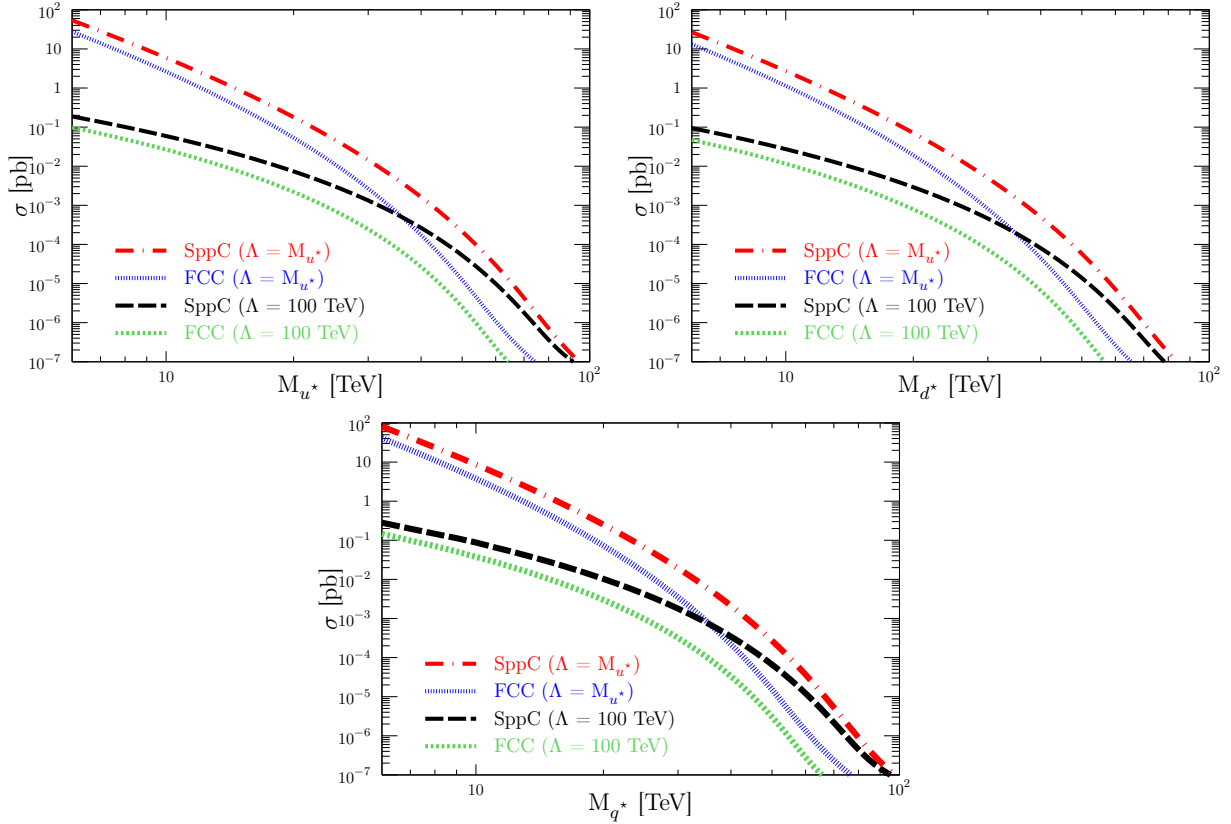

	\begin{center}
		\scalebox{0.43}{\input{pp_jj_cs_100_LEM_ustar.tex}}
		\scalebox{0.43}{\input{pp_jj_cs_100_LEM_dstar.tex}}\\
		\scalebox{0.43}{\input{pp_jj_100_LEM_UD_CS.tex}}
	\end{center}	
	\caption{\label{fig:crossSection}Cross section values of the first generation $u^{\star}$, $d^{\star}$ and $q^{\star}$ (degenerate state) excited quarks at the FCC and SppC.}
	
\end{figure}

In Fig. \ref{fig:crossSection}, first generation excited quarks cross section
values for three cases mentioned above are plotted for the FCC ($\sqrt{s}=100$ TeV) and the SppC
($\sqrt{s}=136$ TeV) with $\Lambda=M_{u^{\star}}$, $\Lambda=M_{d^{\star}}$, $\Lambda=M_{q^{\star}}$ (degenerate state)  and $\Lambda=100$ TeV. When the compositeness scale value is taken as equal to 
excited quark masses, cross section values are about 300 times higher at
6 TeV mass value for both collider options. Indeed, it seems that
excited quark could be produced at very high mass values for both collider
options. It should be noted that as the LHC experimental studies on excited quarks with dijet final states do not consider SM interference contribution to cross section \cite{atlas2017, sirunyan2018}, we did not simulate interference with SM for the FCC at this stage. For the same reason,  QCD corrections were disregarded in this analysis \cite{ atlas2017,sirunyan2018gammajet, cms2011, cmsNOTE2006, harris2011}.  

\section{\label{sec:Signal-BackGround-Analysis}Signal \protect\lowercase{and} Background Analysis}

\begin{figure}[H]
	\begin{center}
		\includegraphics[scale=0.6]{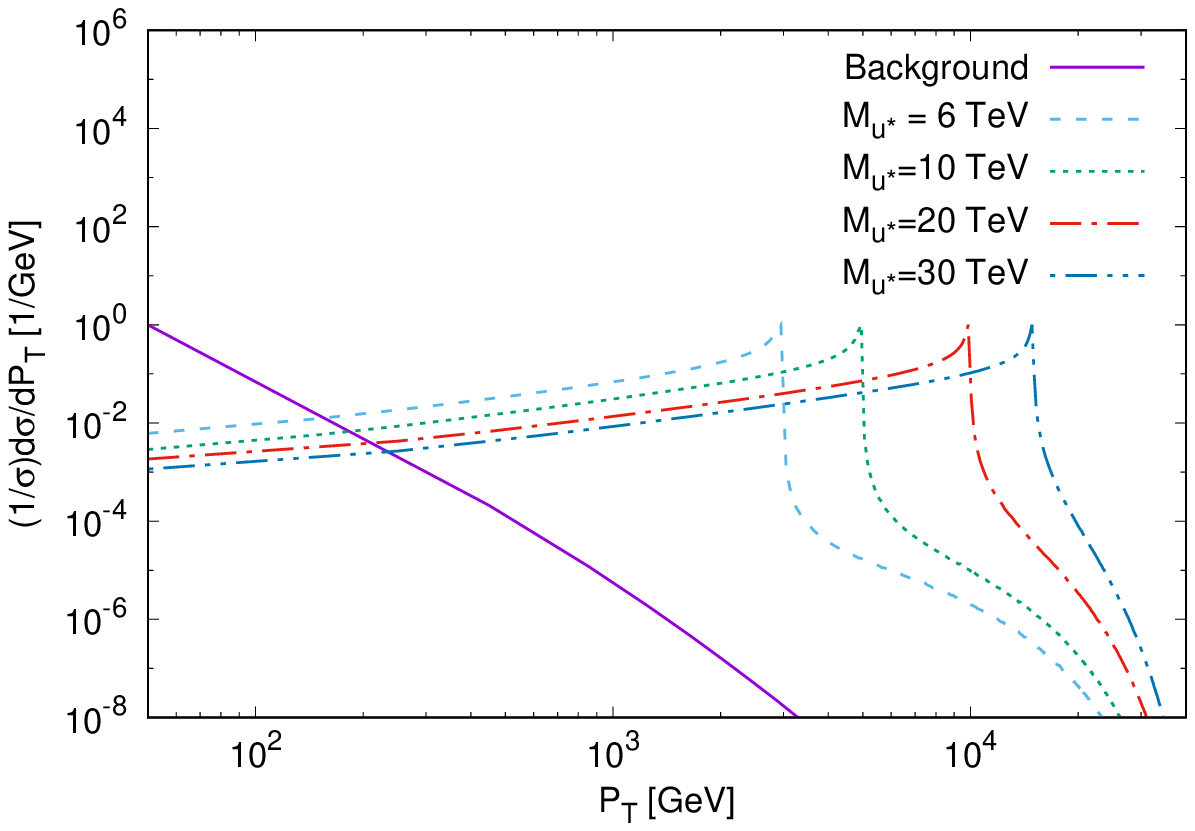}\includegraphics[scale=0.6]{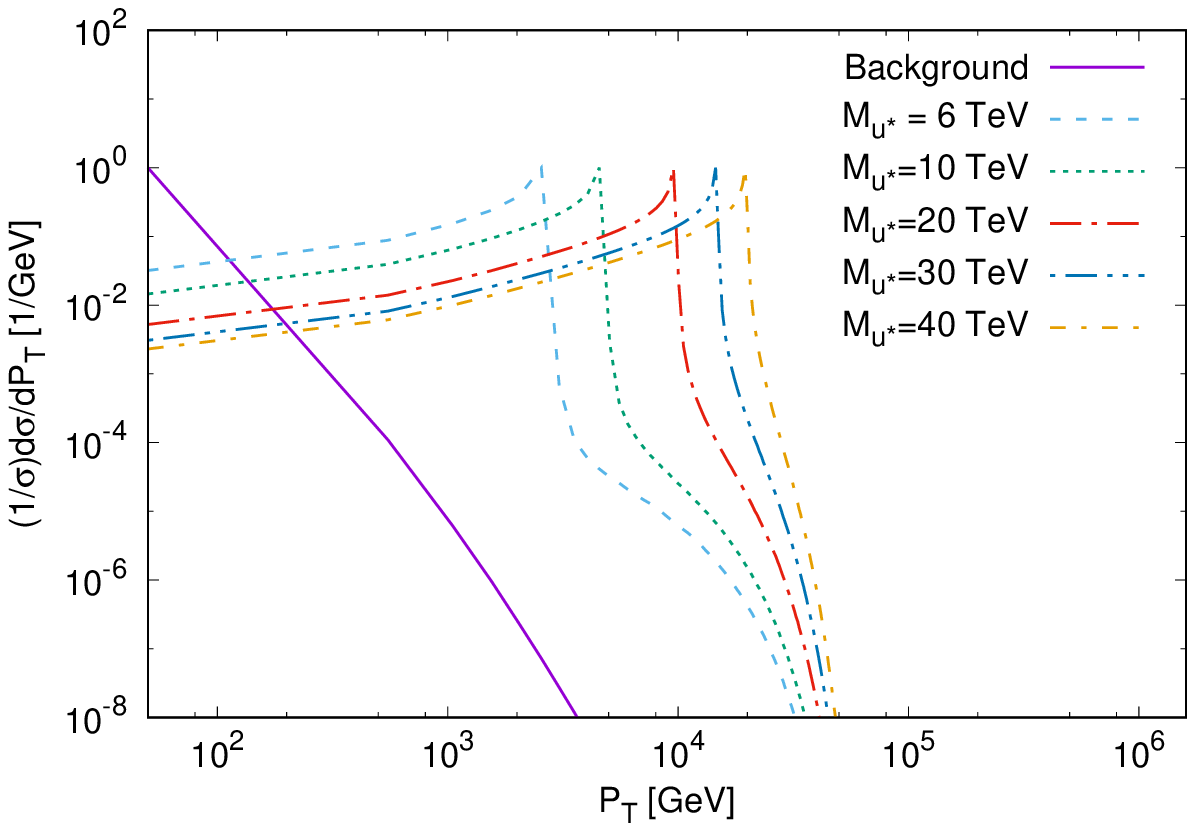}
		\includegraphics[scale=0.6]{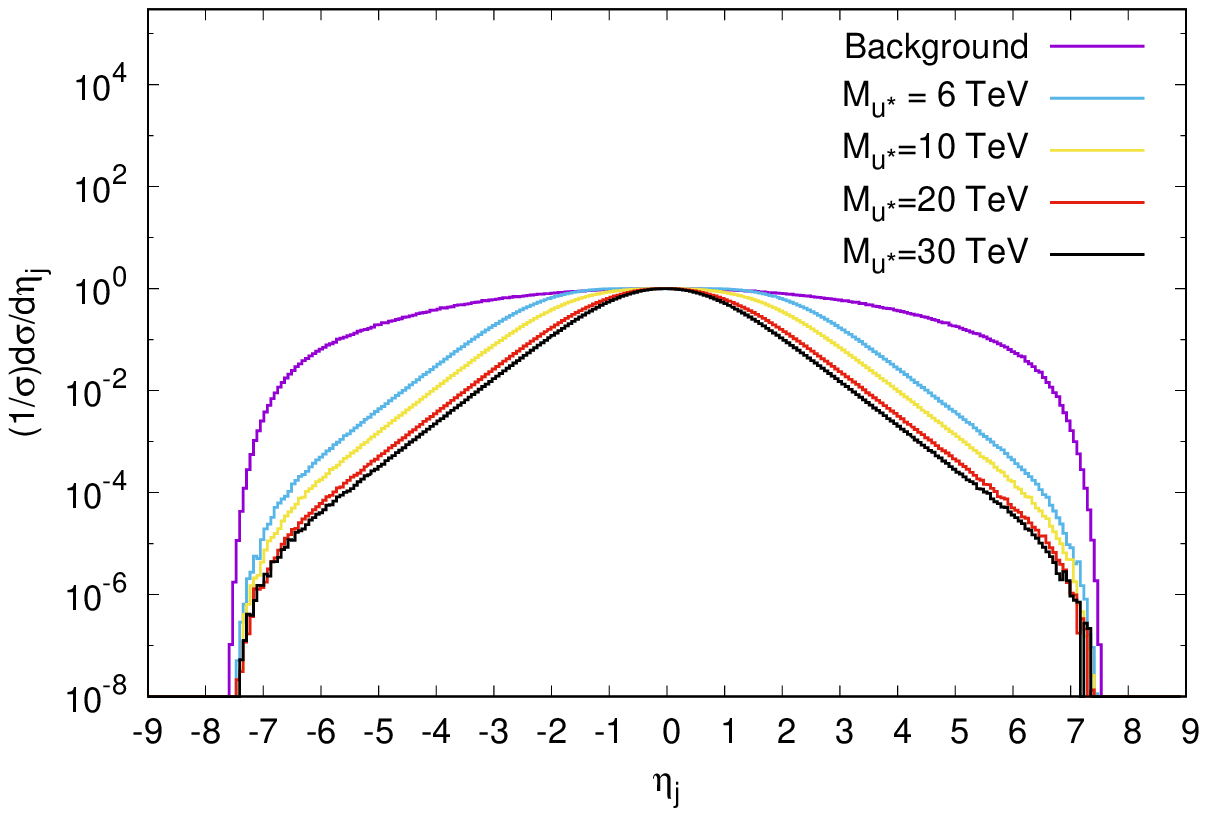}\includegraphics[scale=0.6]{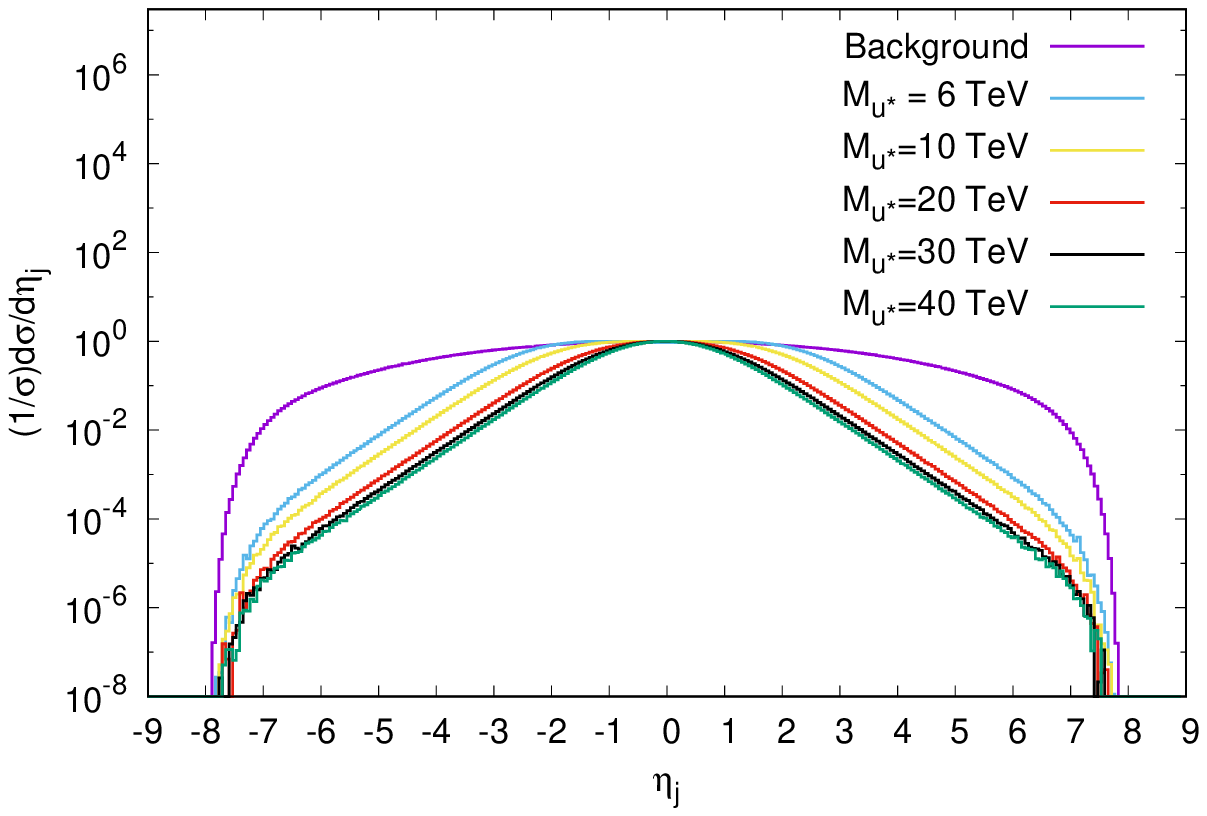}
	\end{center}
	\caption{\label{fig:ptEtaFCC=000026SppC}Transverse momentum and $\eta$
		distribution plots for FCC (left column) and SppC (right column).}
\end{figure}

Signal processes were defined in previous section. Background process which is used in calculation is $pp\rightarrow jj+X$,
here $j$ denotes $u, \, \bar{u},\, d,\,\bar{d},\, c,\, \bar{c},\, s, \,\bar{s},\, b,\, \bar{b}$
and $g$ for three signal cases. It is important to determine transverse momentum ($P_{T}$),
pseudo rapidity ($\eta$) and invariant mass ($M_{jj}$) cut values
for selecting clear signal. To illustrate cut selection, only final state particles distribution originated by excited $u$ quark 
plots are included in Fig. \ref{fig:ptEtaFCC=000026SppC}. According
these figures, $P_{T}$ cuts are applied as 2 TeV, $\eta$ cuts are
determined as $|\eta|<2.5$ in signal and background cross section calculations for three cases. Also, the cone angle radius is chosen as $\Delta R>0.5$ for both colliders. Additionally, invariant mass cuts are applied
as 
$M^{\star}-2\Gamma^{\star}<M_{jj}<M^{\star}+2\Gamma^{\star}$ mass
window for again both collider options, here $M^{\star}$ denotes
excited quarks ($u^{\star}$, $d^{\star}$ and $q^{\star}$) mass and $\Gamma^{\star}$ is total decay widths
of the excited quarks. 

In order to calculate statistical significance, Eq. \ref{eq:significance} is used;

\begin{equation}
SS=\frac{\sigma_{S}}{\sqrt{\sigma_{S}+\sigma_{B}}}\sqrt{\mathcal{L}_{int}}\label{eq:significance}
\end{equation}

where, $\sigma_{S}$ and $\sigma_{B}$ denote signal and background
cross section values, respectively and $\mathcal{L}_{int}$ represents
integrated luminosity. Using Eq. \ref{eq:significance}, we have calculated
excited quarks mass' discovery ($5\sigma$), observation ($3\sigma$) and
exclusion ($2\sigma$) limits on prospective frontier machines, namely,
FCC and SppC. 

\section{\label{sec:Results-and-Conclusions}Results \protect\lowercase{and} Conclusions}

\begin{figure}[H]
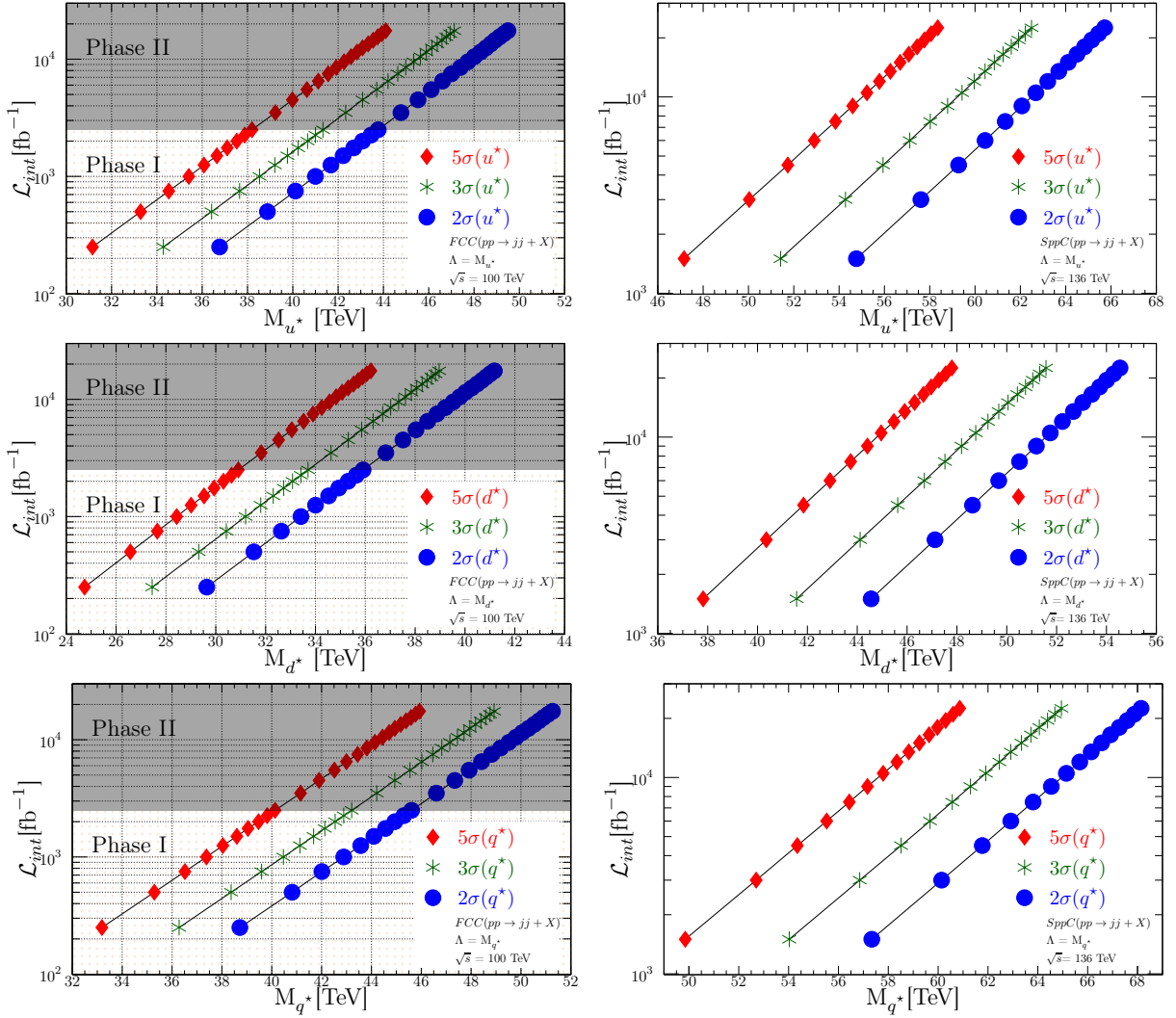

	\begin{center}
		\scalebox{0.43}{\input{FCC_LEM_jj_LumiMassAll_ustar.tex}}
		\scalebox{0.43}{\input{SppC_LEM_jj_LumiMass_ustar.tex}}\\
		\scalebox{0.43}{\input{FCC_LEM_jj_LumiMassAll_dstar.tex}}
		\scalebox{0.43}{\input{SppC_jj_LEM_LumiMass_dstar.tex}}\\
		\scalebox{0.43}{\input{FCC_LEM_jj_LumiMassAll_UD.tex}}
		\scalebox{0.43}{\input{SppC_LEM_jj_LumiMass_ud.tex}}
	\end{center}
	\caption{\label{fig:lumiMass}Mass dependence on luminosity at all confidence levels for the FCC (left column) and SppC (right column). }
\end{figure}

Discovery, observation and exclusion limits on the mass of excited quarks for three cases depending on integrated luminosity of the FCC and SppC with $\Lambda=M^{\star}$ cases are plotted in Fig. \ref{fig:lumiMass}.
Attainable mass limits for all three cases for FCC- Phase I and II, and SppC with their final integrated luminosity values at the end of operating times are listed in Tab. \ref{tab:lumiMass}.  It is seen that FCC-Phase I  will afford an opportunity to discover, observe or exclude, degenerate case of excited quarks up to  40.1, 43.2 and 45.6 TeV  respectively. At the end of the FCC-Phase II, these values become
$M_{q^{\star}}=45.9$ TeV ($5\sigma$), $M_{q^{\star}}=48.9$ TeV ($3\sigma$)
and $M_{q^{\star}}=51.3$ TeV ($2\sigma$). On the other hand, corresponding
values for SppC are $M_{q^{\star}}=60.9$ TeV ($5\sigma$), $M_{q^{\star}}=65.0$
TeV ($3\sigma$) and $M_{q^{\star}}=68.1$ TeV ($2\sigma$) that essentially
exceed the FCC limits. 

\begin{table}[H]
	\caption{Attainable mass limits for all three cases at FCC and SppC with corresponding final integrated luminosity values. Compositeness scale is chosen equal to excited quarks mass values. \label{tab:lumiMass}}
	\resizebox{\textwidth}{!}{
	\begin{tabular}{|l|c|c|c|c|c|c|c|c|c|c|}
		\hline 
		\multicolumn{2}{|p{5cm}|}{\textbf{Colliders}}&  \multicolumn{3}{c|}{\textbf{FCC-PhaseI}} &  \multicolumn{3}{c|}{\textbf{FCC-PhaseII}} & \multicolumn{3}{c|}{\textbf{SppC}} \\ 
		\hline 
		\multicolumn{2}{|l|}{\textbf{Integrated Luminosity [fb$^{-1}$]}}&  \multicolumn{3}{c|}{\textbf{2500}} &  \multicolumn{3}{c|}{\textbf{17500}} & \multicolumn{3}{c|}{\textbf{22500}} \\ 
		\hline 
		\multicolumn{2}{|l|}{\textbf{Significance}} & \textbf{$5\sigma$} &\textbf{$3\sigma$ }& \textbf{$2\sigma$}& \textbf{$5\sigma$} &\textbf{$3\sigma$ }& \textbf{$2\sigma$}&\textbf{$5\sigma$} &\textbf{$3\sigma$ }& \textbf{$2\sigma$} \\ 
		\hline 
		\multirow{3}{5cm}{\textbf{Excited Quark\\ Mass }[TeV]}& \textbf{M$_{u^{\star}}$} & 38.2 & 41.3& 43.8 & 44.1 & 47.1 & 49.5 &58.4  & 62.5  & 65.7 \\ 
		\cline{2-11}
		& \textbf{M$_{d^{\star}}$} & 30.9& 33.7 & 35.9 & 36.3  & 39.0 & 41.2 & 47.8  & 51.6  & 54.5 \\ 
		\cline{2-11}
		& \textbf{M$_{q^{\star}}$}   & 40.1  & 43.2 & 45.6   & 45.9  & 48.9  & 51.3  & 60.9  & 65.0 & 68.1 \\ 
		\hline 
	\end{tabular} 
}
\end{table}
As mentioned above, we did not anticipate  interference of the signal model with the SM contribution. In order to estimate this contribution, we compared discovery limits for $u^{\star}$  at the FCC Phase II. As seen from the Tab. \ref{tab:lumiMass}, this limit was 44.1 TeV in our case. If one takes interference terms into account, discovery  limit becomes 45.0 TeV. The latter value was obtained using same discovery cuts together with corresponding statistical signification equation, namely;

\begin{equation}
SS=\frac{|\sigma_{tot}-\sigma_{B}|}{\sqrt{\sigma_{tot}}}\sqrt{\mathcal{L}_{int}}\label{eq:significance2}
\end{equation}
  
where $\sigma_{tot}$   includes signal, SM and interference contributions.  Interference terms leads to slightly higher discovery limit. Therefore, presented results in this study can be considered as a bit conservative. 

Concerning the role of systematic uncertainties caused by choice of PDF, factorization and renormalization scales, analysis performed at the ATLAS and CMS experiments show that their impact is less than 1\% for $q^{\star}\rightarrow jj$ channel \cite{atlas2017}.  As for the efficiency of jet registration, it is nearly 100\% for jets with $P_T$  above 20 GeV \cite{atlas2017}.
 
\begin{figure}[H]
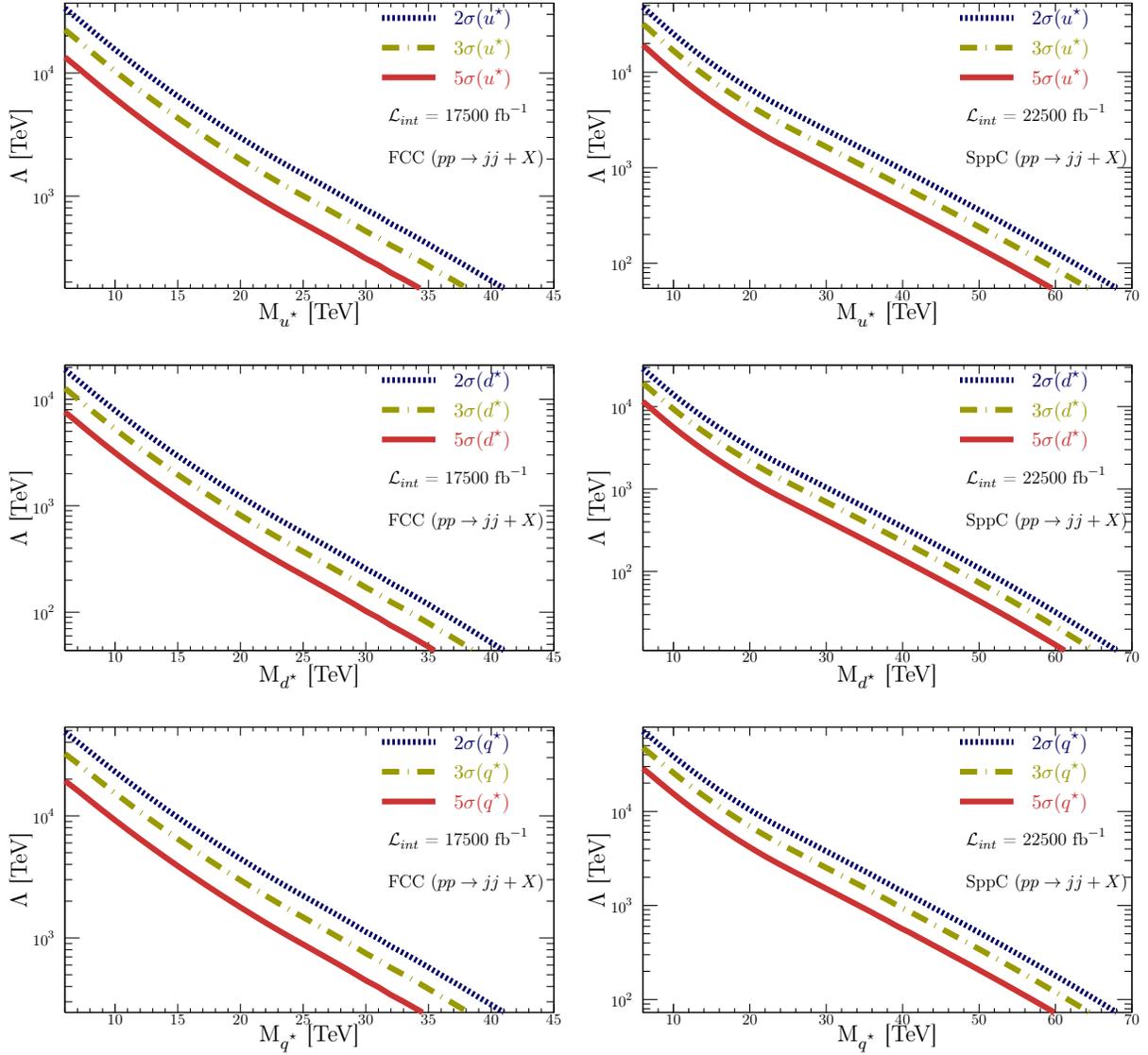

	\begin{center}
		\scalebox{0.43}{\input{ustar_LamdaSearch_FCC_Lumi_17500.tex}}
		\scalebox{0.43}{\input{ustar_LamdaSearch_SppC_Lumi_22500.tex}}\\
		\scalebox{0.43}{\input{dstar_LamdaSearch_FCC_Lumi_17500.tex}}
		\scalebox{0.43}{\input{dstar_LamdaSearch_SppC_Lumi_22500.tex}}\\
		\scalebox{0.43}{\input{ud_LamdaSearch_FCC_Lumi_17500.tex}}
		\scalebox{0.43}{\input{ud_LamdaSearch_SppC_Lumi_22500.tex}}
	\end{center}
	\caption{\label{fig:LamdaSearch}Compositeness scale dependence on $u^{\star}$ mass for the FCC (left column) and SppC (right column)}
\end{figure}

In principle, compositeness scale might be quite higher than excited
quark mass. If excited $u$ and $d$ quarks are not discovered at FCC or SppC,
one can evaluate lower limits on compositeness scale. For illustration,
we plot achievable compositeness scale depending on $u^{\star}$, $d^{\star}$ and $q^{\star}$ mass
for ultimate luminosity values at both colliders in Fig. \ref{fig:LamdaSearch}. If it is assumed that $u^{\star}$ mass equals 20 TeV and it is not seen at FCC in resonant channel, according to Fig. \ref{fig:LamdaSearch}, this means that compositeness scale is larger than 1.2 PeV ($5\sigma$), 2.0 PeV ($3\sigma$) and 3.0 PeV ($2\sigma$). Achievable scales for other values of $u^{\star}$  as well as $d^{\star}$ and $q^{\star}$ (degenerate state) are presented in Tab. \ref{tab:LamdaSearchFCC}. Similar results for the SppC are given in Tab. \ref{tab:LamdaSearchSppC}. 

\begin{table}[H]
	\caption{Compositeness scale values corresponding to some selected mass quantities for all three cases at FCC with final integrated luminosity values. \label{tab:LamdaSearchFCC}}
		\normalsize\setlength{\tabcolsep}{12pt}
		\begin{tabular}{l@{\hspace{12pt}} *{11}{c|}}	
			\hline 
			\multicolumn{11}{|c|}{\textbf{FCC (\text{$\mathcal{L}_{int}$}=17500 fb$^{-1}$)}}   \\ \hline 		 
			\multicolumn{2}{|c|}{\multirow{3}{1cm}{\textbf{Mass [TeV]}}} & \multicolumn{9}{c|}{\textbf{Compositeness Scale $\Lambda$} [PeV]} \\ \cline{3-11}
				\multicolumn{2}{|c|}{ }& \multicolumn{3}{c|}{\textbf{$u^{\star}$}} &   \multicolumn{3}{c|}{\textbf{$d^{\star}$}}   &   \multicolumn{3}{c|}{\textbf{$q^{\star}$}} \\ \cline{3-11}
		\multicolumn{2}{|c|}{ }& \textbf{$5\sigma$ }&\textbf{$3\sigma$ } &\textbf{$2\sigma$ } &\textbf{$5\sigma$ } &\textbf{$3\sigma$ } &\textbf{$2\sigma$}  & \textbf{$5\sigma$} & \textbf{$3\sigma$} & \textbf{$2\sigma$} \\ \hline
		 \multicolumn{2}{|c|}{\textbf{6 }}& 13.5  &  22.4 & 33.6  & 7.61 & 12.7 & 19.0 & 19.4 &       32.3  &       48.5\\	\cline{2-11}
		\multicolumn{2}{|c|}{\textbf{10}} & 6.21 &     10.4  &     15.5    & 3.15 &      5.25  &      7.88  & 9.22  &      15.4  &      23.1\\ \cline{2-11}
	\multicolumn{2}{|c|}{\textbf{20}} & 1.20       & 1.99    & 2.99 & .489  &      .815 &      1.22 & 1.78  &     2.97  &       4.46\\	\cline{2-11}
		\multicolumn{2}{|c|}{\textbf{30}}& .311  &  .518      &  .776 & .102  &   .171  &  .256  & .448  &     .747 &   1.12\\
			\hline 
				
		\end{tabular} 
\end{table}

\begin{table}[H]
	\caption{Compositeness scale values corresponding to some selected mass quantities for all three cases at SppC  with final integrated luminosity values. \label{tab:LamdaSearchSppC}}
			\normalsize\setlength{\tabcolsep}{12pt}
		\begin{tabular}{l@{\hspace{12pt}} *{11}{c|}}
			\hline 
			\multicolumn{11}{|c|}{\textbf{SppC (\text{$\mathcal{L}_{int}$}=22500 fb$^{-1}$)}}   \\ \hline	
			\multicolumn{2}{|c|}{\multirow{3}{1cm}{\textbf{Mass [TeV]}}} & \multicolumn{9}{c|}{\textbf{Compositeness Scale $\Lambda$} [PeV]} \\ \cline{3-11}	
			\multicolumn{2}{|c|}{ }& \multicolumn{3}{c|}{\textbf{$u^{\star}$}} &   \multicolumn{3}{c|}{\textbf{$d^{\star}$}}   &   \multicolumn{3}{c|}{\textbf{$q^{\star}$}} \\ \cline{3-11}
			\multicolumn{2}{|c|}{ }& \textbf{$5\sigma$ }&\textbf{$3\sigma$ } &\textbf{$2\sigma$ } &\textbf{$5\sigma$ } &\textbf{$3\sigma$ } &\textbf{$2\sigma$}  & \textbf{$5\sigma$} & \textbf{$3\sigma$} & \textbf{$2\sigma$} \\ \hline
			 \multicolumn{2}{|c|}{\textbf{6 }}& 19.2   &      32.0  &      48.0 & 11.5  &      19.2  &      28.8 & 28.6 &       47.6  &      71.4  \\ \cline{2-11}
		\multicolumn{2}{|c|}{\textbf{10}} & 10.1  &  16.8  & 25.2 & 5.59 & 9.31  & 14.0  & 15.4 & 25.7 & 38.5 \\ \cline{2-11}
		\multicolumn{2}{|c|}{\textbf{20}} & 2.68 &  4.47  & 6.71 & 1.29  &  2.15  & 3.23 & 4.16 & 6.94 & 10.4 \\ \cline{2-11}
		\multicolumn{2}{|c|}{\textbf{30}} & .993 &  1.66  & 2.48 & .418  & .696  &  1.04 & 1.51 & 2.52 & 3.78 \\ \cline{2-11}
		\multicolumn{2}{|c|}{\textbf{40}}& .383 &  .638  & .957 & .139  & .231  &  .347 & .562 & .936 & 1.41 \\ 
			\hline 
		\end{tabular} 
\end{table}

\begin{figure}[H]
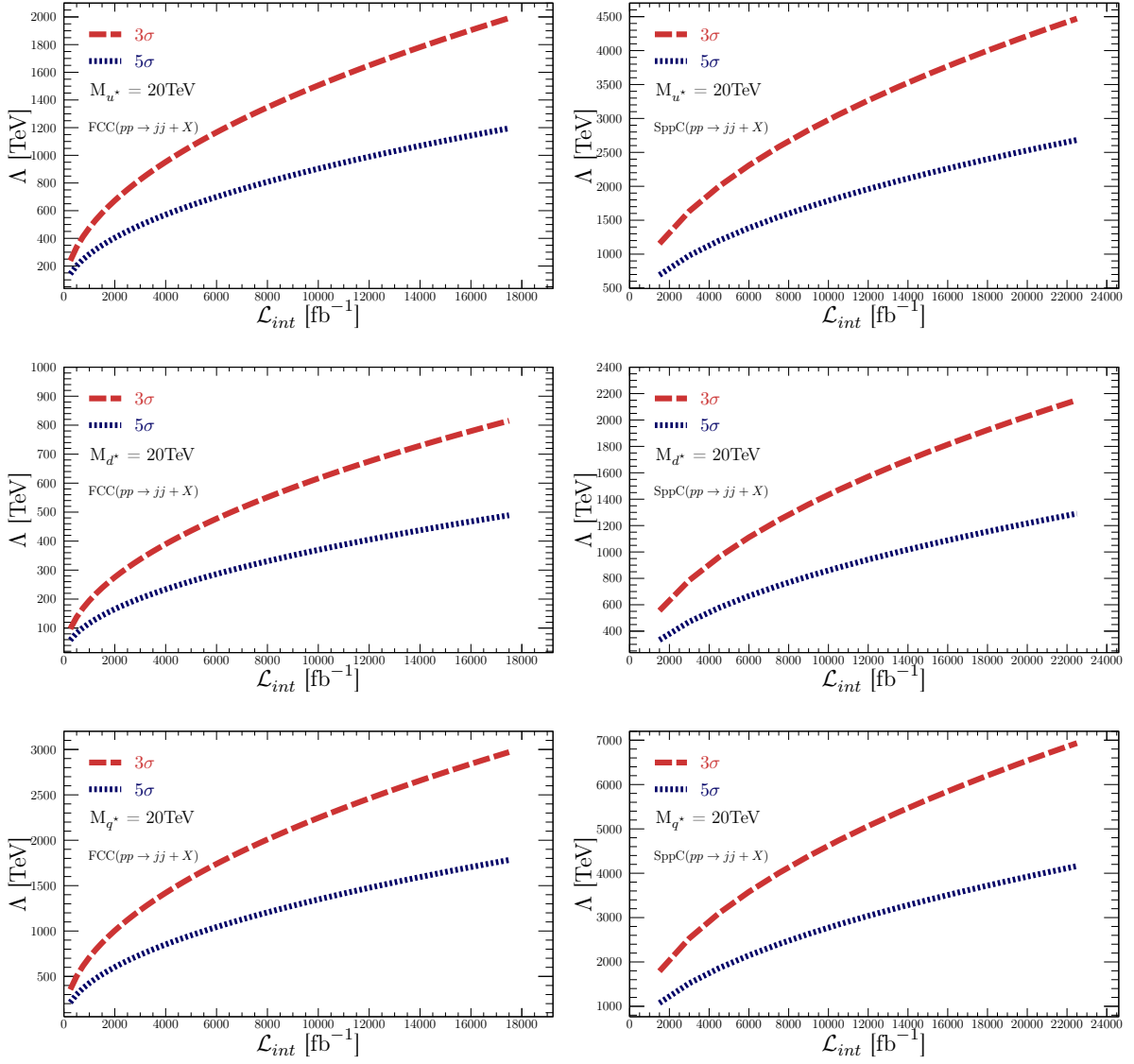

	\begin{center}
		\scalebox{0.43}{\input{ustar_FCC_LVL_Mass20.tex}}
		\scalebox{0.43}{\input{ustar_SppC_LVL_Mass20.tex}}\\
		\scalebox{0.43}{\input{dstar_FCC_LVL_Mass20.tex}}
		\scalebox{0.43}{\input{dstar_SppC_LVL_Mass20.tex}}\\
		\scalebox{0.43}{\input{qstar_FCC_LVL_Mass20.tex}}
		\scalebox{0.43}{\input{qstar_SppC_LVL_Mass20.tex}}
	\end{center}
	
\caption{\label{fig:LamdaLumi}Compositeness scale - luminosity correlation plots for the FCC (left column) and SppC (right column).}
\end{figure}

In Fig. \ref{fig:LamdaLumi}, necessary luminosities for observation
and discovery of 20 TeV mass excited u quark depending on compositeness
scale are plotted for both energy-frontier colliders. It is seen
that if $\Lambda=1000$ TeV, FCC will observe $u^{\star}$ with $4500\;fb^{-1}$
integrated luminosity and $\mathcal{L}_{int}=12000\:fb^{-1}$is needed
for discovery, which correspond to 12 and 19.5 operation years, respectively.
Concerning the SppC, it will observe $u^{\star}$ with 20 TeV mass within
first year and discover it in 2 years if compositeness scale is equal
to 1000 TeV. 

In conclusion, FCC and SppC have excellent potential for discovery of excited $u$ and $d$ quarks. If compositeness scale coincide with excited quark masses , FCC reach M$_{u^{\star}}$ = 44 TeV , M$_{d^{\star}}$ = 36 TeV and M$_{q^{\star}}$ = 46 TeV (degenerate state). Corresponding values for SppC are 58 TeV, 48 TeV, and 61 TeV, respectively. If compositeness scale is higher than excited quark masses, discovery of excited quarks will afford an opportunity to determine $\Lambda$ at the same time.

\begin{acknowledgments}
This study is supported by TUBITAK under the grant No [114F337]. Authors are grateful to the Usak University, Energy, Environment and Sustainability Application and Research Center for their support. 
\end{acknowledgments}

\bibliographystyle{apsrev4-1}
\bibliography{excited_quark}

\end{document}